\documentclass[letterpaper,floatfix,showpacs,aps,prl,twocolumn,10pt]{revtex4}

\usepackage{latexsym,amsmath,amssymb,amsfonts,mathbbol,graphicx,color}

\begin{document}

\newtheorem{theorem}{Theorem}
\newtheorem{definition}{Definition}
\newtheorem{example}{Example}
\newtheorem{corollary}{Corollary}

\def\FCW{0.98\columnwidth}
\def\HPW{0.48\textwidth}
\def\TQPW{0.73\textwidth}
\def\FPW{0.98\textwidth}
\def\Id{1\!\mathrm{l}}
\newcommand{\qed}{\hfill $\Box$ \hfill \\}

\title{The structure of preserved information in quantum processes}
\author{Robin Blume-Kohout$^1$, Hui Khoon Ng$^1$, David Poulin$^2$, 
and Lorenza Viola$^3$}
\affiliation{$^1$Institute for Quantum Information, 
$^2$Center for the Physics of Information, Caltech, Pasadena, CA 91125, USA \\
\mbox{$^3$Department of Physics and Astronomy, 
Dartmouth College, 6127 Wilder Laboratory, Hanover, NH 03755, USA}}

\date{\today}
\begin{abstract}
We introduce a general operational characterization of
information-preserving structures (IPS) -- encompassing noiseless
subsystems, decoherence-free subspaces, pointer bases, and
error-correcting codes -- by demonstrating that they are isometric to
fixed points of unital quantum processes. Using this, we show that
every IPS is a matrix algebra. We further establish a structure
theorem for the fixed states and observables of an arbitrary process,
which unifies the Schr\"odinger and Heisenberg pictures, places
restrictions on physically allowed kinds of information, and provides
an efficient algorithm for finding all noiseless and unitarily
noiseless subsystems of the process.
\end{abstract}


\maketitle

Quantum processes -- a.k.a. quantum channels, quantum operations, or
completely positive (CP) maps -- \cite{KrausBook83,NielsenBook00} --
are central to the theory and practice of quantum information
processing (QIP).  They describe how quantum states evolve over a
period of time in the presence of noise, or how a device's output
depends on its input.  They are also complex and unwieldy: to fully
specify a quantum process on a $d$-dimensional system requires $d^4$
real numbers.  Most of this data is irrelevant to what one really
wants to know: What information can pass unharmed through the process?
Beside being central to QIP, a general answer is broadly relevant to
both fundamental physics and quantum technologies, for the
information-preserving degrees of freedom are precisely those that may
be reliably characterized and exploited.  \emph{Information-preserving
structures} (\textbf{IPS}) in quantum processes -- what they are and
how to find them -- are the subject of this Letter.

The quest for such structures has a long history in quantum physics.
\emph{Pointer states} (PS), defined in the context of quantum
measurement theory, are ``most classical'' states that resist
decoherence \cite{Zurek}.  QIP science has spurred interest in the
preservation of \emph{quantum} information, leading to the notion of
\emph{noiseless subsystems} (NS) \cite{NS} as passive IPS that emerge
from the existence of symmetries in the noise, and recover both
\emph{decoherence-free subspaces} (DFS) \cite{DFS}, and PS in special
limits.  Processes admitting no NS may still preserve information,
which can be actively protected using \emph{quantum error correction}
(QEC) \cite{QEC,KLP05a} to create an effective NS.  Rapid experimental
progress in implementing DFS \cite{DFSexp}, NS \cite{NSexp}, and QEC
\cite{QECexp} heightens the need for a complete and constructive
characterization of preserved information.

In this Letter, we formulate a general {\em operational} theory of
IPS.  The key insight is to identify preserved information with sets
of states (or \emph{codes}) whose \emph{mutual distinguishability} is
left unchanged.  We prove that every preserved code can, through error
correction, be made \emph{noiseless}, then show that every optimal
noiseless code is isometric \cite{iso} to the fixed-point set of the
dynamics.  This set, in turn, is isometric to a matrix algebra, thus
we conclude that every IPS is an algebra.  Finally, we provide an
explicit structure for the fixed points of an arbitrary process, and
an efficient algorithm to determine its noiseless and unitarily
noiseless IPS.

Our results fill several gaps in existing work.  Starting from basic
operational definitions, our approach encompasses everything that
could represent information perfectly preserved by a quantum process, and shows
an explicit connection to fixed points.  Our structure theorem for
fixed points is general, whereas previous results applied only to
unital \cite{AGG02a,KribsPEMS03} maps, or ones with a full-rank fixed
state \cite{Frigerio}.  While information preservation has been
addressed in both the Schr\"odinger and Heisenberg \cite{Beny}
dynamical pictures, we consistently unify them.  Available algorithms
to find IPS are either inefficient (e.g., Zurek's ``predictability
sieve'' for PS \cite{ZurekPTP93}, or Choi and Kribs's method for NS
\cite{ChoiPRL06}), or restricted to purely noiseless information
\cite{Knill06a} or unital channels \cite{KS06a}.  By explicitly
shifting focus from the noise commutant to the fixed-point set (recent
work, e.g.  \cite{Beny}, has also moved in this direction) our
approach paves the way to analyzing ``approximate'' IPS, beyond
existing results on the stability of DFS/NS under symmetry-breaking
perturbations \cite{BaconPRA99}.

\textbf{Quantum states and processes:} We consider an open quantum
system with a [finite] $d$-dimensional Hilbert space $\mathcal{H}$.
Its state is described by a non-negative, trace-1, $d\times d$ density
matrix $\rho$, which is also a vector in the system's {Hilbert-Schmidt
space} $\mathcal{B(H)}$ (the space of bounded operators on
$\mathcal{H}$).  The system's dynamical evolution over time $t$ is
described by a \emph{quantum process}
$\mathcal{E}:\mathcal{B(H)}\rightarrow\mathcal{B(H)}$.  $\mathcal{E}$
is linear, trace-preserving (TP), and CP, which ensures that
$\mathcal{E}$ does not produce negative probabilities operating on
arbitrary states.  $\mathcal{E}$ is CP iff $\mathcal{E}(\rho) =
\sum_i{K_i\rho K_i^\dagger}$ for some set of \emph{Kraus operators}
$\{K_i\}$, and TP iff $\sum_i{K_i^\dagger K_i}=\Id$. $\mathcal{E}$ is
{\em unital} iff, in addition, $\mathcal{E}(\Id)= \sum_i{K_i
K_i^\dagger}=\Id$ (see \cite{KrausBook83,NielsenBook00} for further
details).

\textbf{Preserved information and distinguishability:} To encode
information, we prepare the system in a state $\rho$, chosen from a set
$\mathcal{C}$ of possible states.  We denote any such $\mathcal{C}$ a
\emph{code}, without {\em a priori} assuming any structure for
$\mathcal{C}$.  The code defines the kind of information encoded; in
particular, our definition includes all the familiar examples: e.g., a
QEC code contains all the states in a subspace $\mathcal{P}\subseteq
\mathcal{H}$; a classical code comprises a discrete set of orthogonal
states.  Many other kinds of codes are possible, and our first goal is
to classify them.

To access the information, we must \emph{distinguish} between states
$\rho,\rho' \in\mathcal{C}$.  If we assign prior probabilities
$\{q,1-q\}$ to $\rho$ and $\rho'$, and make the optimal measurement to
distinguish them, we guess correctly with probability $p =
\frac12\left(1+\left\|q\rho-(1-q)\rho'\right\|_1\right)$ (see
Ref. \cite{HelstromBook76}, IV.2).  Clearly, if $\mathcal{E}$ makes
the states in $\mathcal{C}$ less distinguishable, then information was
not perfectly preserved.  We therefore propose the following
operational criterion: \emph{A code $\mathcal{C}$ is preserved by a
process $\mathcal{E}$ iff each pair of states
$\rho,\rho'\in\mathcal{C}$ is just as distinguishable \emph{after}
$\mathcal{E}$ as before it.}  The distinguishability result cited
above implies then a technical definition: \emph{$\mathcal{C}$ is
preserved by $\mathcal{E}$ iff, for every $\rho,\rho'\in\mathcal{C}$
and $x\in\mathbb{R}_+$, $\left\|\mathcal{E}(\rho-x\rho')\right\|_1 =
\left\|\rho-x\rho'\right\|_1$.}

A useful consequence is that preserved codes can always be closed
under (real) linear combination, so we can assume that $\mathcal{C}$
comprises all the states in an \emph{operator subspace} of
$\mathcal{B(H)}$.  $\mathcal{C}$ is preserved if it is
\emph{isometric} to $\mathcal{E(C)}$, that is, $\mathcal{E}$ acts as a
1:1 trace-distance-preserving map on $\mathcal{C}$.  Several
operational notions of ``preserved'' will be relevant. From strongest
to weakest:

1. $\mathcal{C}$ is \textbf{noiseless} for $\mathcal{E}$ iff it is
preserved by any convex mixture
$\sum_{n}{p_n\mathcal{E}^n}$, with $p_n\geq 0$ and $\sum_n{p_n}=1$.

2. $\mathcal{C}$ is \textbf{unitarily noiseless}~\cite{unins} for
$\mathcal{E}$ iff it is preserved by $\mathcal{E}^n$, for every fixed
$n\in\mathbb{N}$.

3. $\mathcal{C}$ is \textbf{correctable} for $\mathcal{E}$ iff there
exists a correction process $\mathcal{R}$ such that $\mathcal{C}$ is
noiseless for $\mathcal{R\circ\mathcal{E}}$.

Thus, while both noiseless and unitarily noiseless codes preserve
information indefinitely without any intervention, they differ in how
the preserved information ``moves around". The optimal measurement to
distinguish two states in a noiseless code is independent of the
number $n$ of applications of $\mathcal{E}$ (and can be derived from
$\mathcal{E}_\infty$, see proof of Theorem
\ref{thmNoiselessFixed}). In contrast, for a unitarily noiseless code
(e.g., a system evolving unitarily, as $\mathcal{E}(\rho) = U\rho
U^\dagger$) this measurement may depend on $n$, so we must keep track
of how many times $\mathcal{E}$ has occurred.  Correctable codes, on
the other hand, are not inherently stable -- but they can be
stabilized indefinitely by applying $\mathcal{R}$.  We can collapse
the lowest levels of the above hierarchy:
\begin{theorem}  
A code $\mathcal{C}$ is preserved by the process $\mathcal{E}$ iff it
is correctable for $\mathcal{E}$. \label{thmPreservedIsCorrectable}
\end{theorem}
While we defer a full proof to Ref. \cite{RBK07c}, the central idea is
simple: If $\mathcal{C}$ is preserved, then we can correct it with the
\emph{transpose channel} \cite{Barnum00},
$$\widehat{\mathcal{E}}_\mathcal{P}(\rho) = \sum_i{\left(PK_i^\dagger
P\mathcal{E}(P)^{-1/2}\right)\rho\left(\mathcal{E}(P)^{-1/2} P K_i
P\right)},
$$ 
\noindent 
where $\mathcal{P}$ is the joint support of every
$\rho\in\mathcal{C}$, and $P$ projects onto it. 
Notice that $\widehat{\mathcal{E}}_{\mathcal P}\circ\mathcal{E}(P) =
\mathcal{E}^\dagger(\mathcal{E}(P)^{-1/2}\mathcal{E}(P)\mathcal{E}(P)^{-1/2})
= P,$ thus the corrected map is not only TP but also unital on the
code's support.

Because a correctable code for $\mathcal{E}$ is a noiseless code for
some other channel $\mathcal{R}\circ\mathcal{E}$, we can characterize
all preserved codes by characterizing noiseless codes.  The first step
is to relate $\mathcal{E}$'s noiseless codes to its fixed points:
\begin{theorem} If $\mathcal{C}$ is a noiseless code for $\mathcal{E}$, then
$\mathcal{C}$ is isometric to a subset of the fixed states of
$\mathcal{E}$. \label{thmNoiselessFixed}
\end{theorem}
\textbf{Proof}: $\mathcal{C}$ is preserved by any channel of the form
$\sum_n{p_n\mathcal{E}^n}$ ($\sum{p_n}=1$), including $\mathcal{E}_N =
\frac{1}{N+1}\sum_{n=0}^{N}{\mathcal{E}^n}$, and therefore also by
$\mathcal{E}_\infty = \lim_{N\rightarrow\infty}{\mathcal{E}_N}$
\cite{Lin99a} (the limit is well-defined for finite-dimensional
$\mathcal{H}$).  Thus, $\mathcal{C}$ is isometric to
$\mathcal{E}_\infty(\mathcal{C})$.  But
$\mathcal{E}\circ\mathcal{E}_\infty = \mathcal{E}_\infty$, so if
$\sigma = \mathcal{E}_\infty(\rho)$, then $\mathcal{E}(\sigma) =
\sigma$.  Therefore, $\mathcal{E}_\infty$ projects onto the fixed
points of $\mathcal{E}$, so $\mathcal{E}_\infty(\mathcal{C})$ is a
subset of $\mathcal{E}$'s fixed states.  \hfill\qed

Theorem \ref{thmNoiselessFixed} has two important consequences for
\emph{optimal} codes -- ones that encode as many states as possible.
First, \emph{every optimal noiseless code for $\mathcal{E}$ is
isometric to the set of all fixed states of $\mathcal{E}$.}  The fixed
states are themselves a noiseless code $\mathcal{C}_0$, so if
$\mathcal{C}$ is not isometric to $\mathcal{C}_0$, then it is
isometric to a proper subset, and cannot be optimal.  Next,
\emph{every optimal preserved code for $\mathcal{E}$ is isometric to
the set of all fixed states of a unital, TP map}.  This follows from
Theorem \ref{thmPreservedIsCorrectable}.  If $\mathcal{C}$ is
preserved, then $\widehat{\mathcal{E}}_{\mathcal P}$ corrects it, so
$\mathcal{C}$ is noiseless for $\widehat{\mathcal{E}}_{\mathcal
P}\circ\mathcal{E}$, and (by Theorem \ref{thmNoiselessFixed}),
isometric to its fixed points.  Optimal preserved codes come in
equivalence classes characterized by fixed geometries (the pairwise
distances between elements of a set define its geometry): $\mathcal C$
and $\mathcal{C}'$ are equivalent iff they are isometric.  Equivalent
codes use different states to encode the same information -- they are
manifestations of the same IPS:

\vspace*{2mm}

\noindent
{\bf Definition 1} {\em An IPS of a process $\mathcal{E}$ is the
geometric structure common to an equivalence class of optimal
preserved codes.}

\vspace*{2mm}

An optimal preserved code is isometric to the fixed-point set of
$\widehat{\mathcal{E}}_{\mathcal P} \circ \mathcal{E}$.  Because this
set (and its geometry) depend on $\mathcal P$, $\mathcal E$ may have
several distinct IPS. However, all its optimal {\em noiseless} codes
belong to a single class as they all share the geometry of $\cal E$'s
fixed-point set.  They are manifestations of a unique noiseless IPS:

\vspace*{2mm}

\noindent
{\bf Definition 2} {\em The noiseless IPS of a process $\mathcal{E}$
is the unique geometric structure common to all of its optimal
noiseless codes.}

\vspace*{2mm}

\textbf{The structure of codes:} The next step toward characterizing
the possible IPS is to determine the structure of fixed states for
arbitrary $\mathcal{E}$.  Because $\mathcal{E}$ is linear, its fixed points are
closed under linear combination, hence form an operator subspace of
$\mathcal{B(H)}$. For the special case where $\mathcal{E}$ is
\emph{unital}, several authors have shown \cite{AGG02a,KribsPEMS03}
that: (a) The fixed points of $\mathcal{E}$ form a complex matrix
algebra $\mathcal{A}$; (b) $\mathcal{A}$ is the commutant of
$\mathcal{E}$'s Kraus operators; (c) $\mathcal{E}$ and
$\mathcal{E}^\dagger$ have the same fixed points.

This is a powerful result because finite-dimensional matrix algebras
share an elegant structure: Every such matrix algebra is a direct sum
of the form,
\begin{equation}
\mathcal{A} = \bigoplus_k{\mathcal{M}_{d_k}\otimes\Id_{n_k}}, \;\;
n_k, d_k \in {\mathbb N},
\label{structure}
\end{equation}
where $\mathcal{M}_{d_k}$ is the algebra of all $d_k\times d_k$
matrices, and $\Id_{n_k}$ is the trivial algebra containing the
$n_k$-dimensional identity \cite{Dav96a}.  Thanks to this result, we
have all the ingredients to describe the structure of preserved
information for an arbitrary (not necessarily unital) $\mathcal{E}$:
\emph{Every optimal preserved code is isometric to a matrix algebra.}
This follows from Theorem \ref{thmPreservedIsCorrectable} (preserved
codes are correctable, with $\mathcal{R}\circ\mathcal{E}$ unital) and
Theorem \ref{thmNoiselessFixed} (optimal noiseless codes are isometric
to fixed point sets), together with the structure theorem cited above.
We conclude: \emph{Any IPS of a process on a $d$-dimensional system is
a subalgebra of $\mathcal{M}_d$}.

\textbf{Fixed points of arbitrary maps:} While the above IPS
characterization is fully general, it is non-constructive as long as
the projector $P$ required to construct the transpose map is unknown.
However, on one hand noiseless codes are isometric to the fixed states
of $\mathcal{E}$ itself (rather than $\widehat{\mathcal{E}}_{\mathcal
P}\circ\mathcal{E}$). On the other hand, the set of all fixed states
is an optimal noiseless code, whose unique isometric algebra
$\mathcal{A}$ fully specifies $\mathcal{E}$'s noiseless IPS. To obtain
a \emph{constructive} characterization of this IPS, we need (1) a
general description of the fixed states of $\mathcal{E}$; and (2) a
way to extract the algebra to which they are isometric.
Unfortunately, the structure theorem for unital maps does not extend
to arbitrary processes.  The following example violates every point
listed earlier: $\mathcal{E}$ and $\mathcal{E}^\dagger$ have different
fixed-point sets, which do not form algebras, and do not commute with
the Kraus operators!

\vspace*{2mm}

\noindent 
{\bf Example:} 
Let $A$ be a qutrit and $B$ a qubit, and $\mathcal{E} =
\mathcal{E}_A\otimes\mathcal{E}_B$ be a process on
$\mathcal{H}_A\otimes\mathcal{H}_B$, with Kraus operators
\begin{eqnarray*}
\mathcal{E} \sim \begin{array}{l}
\left\{|0\rangle\!\langle0|+|1\rangle\!\langle1|,
\frac{1}{\sqrt2}|0\rangle\!\langle2|,\frac{1}{\sqrt2}|1\rangle\!\langle2|\right\}_A
\\ \otimes
\left\{\frac12|0\rangle\!\langle0|,\frac12|0\rangle\!\langle1|,
\frac{\sqrt3}{2}|1\rangle\!\langle0|,\frac{\sqrt3}{2}|1\rangle\!\langle1|\right\}_B
\end{array}.
\end{eqnarray*}
$\mathcal{E}$ does nothing to the $\{|0\rangle,|1\rangle\}$ subspace
of $A$, but maps $|2\rangle_A$ into an equal mixture of
$|0\rangle\!\langle0|_A$, $|1\rangle\!\langle1|_A$. At the same time,
it forces $B$ into $\tau_B = \frac14|0\rangle\!\langle0|_B
+\frac34|1\rangle\!\langle1|_B$.  $\mathcal{E}$'s fixed states are
$\sigma_{A} \otimes\tau_B$ (for any $2\times2$ matrix $\sigma_A$), and
the fixed observables of $\mathcal{E}^\dagger$ are
$\left(\sigma_{A}+\frac12\mathrm{Tr}(\sigma_A)
|2\rangle\!\langle2|_A\right)\otimes\Id_B$.  The commutant of the
Kraus operators is nothing but $\Id$.

\vspace*{2mm}

Still, we \emph{can} characterize fixed states and observables:
\begin{theorem}
Let $\mathcal{E}$ be a quantum process on $\mathcal{B(H)}$, $\Sigma$
the fixed points of $\mathcal{E}$, and $\mathcal{B}$ the fixed points
of $\mathcal{E}^\dagger$.  Then:

\emph{(i)} $\Sigma$ and $\mathcal{B}$ are each isometric to a matrix algebra
$\mathcal{A}\subseteq\mathcal{B(P)}$, where $\mathcal{P}$ is a
subspace of $\mathcal{H}$.

\emph{(ii)} $\Sigma$ is supported on $\mathcal{P}$, and contains all
operators $\sigma = \bigoplus_k{M_{d_k}\otimes\tau_{n_k}}$, where
$M_{d_k}$ is an arbitrary $d_k\times d_k$ operator, and $\tau_{n_k}$
is a fixed $n_k\times n_k$ state.

\emph{(iii)} $\mathcal{B}$ contains all operators of the form $X =
A_\mathcal{P} \oplus
\mathcal{F}_{\mathcal{P}\rightarrow\overline{\mathcal{P}}}
\left(A_\mathcal{P}\right)$,
where $A_\mathcal{P}\in\mathcal{A}$, $\overline{\mathcal{P}}$ is the
complement of $\mathcal{P}$ in $\mathcal{H}$, and
$\mathcal{F}_{\mathcal{P}\rightarrow\overline{\mathcal{P}}}$ is a
fixed linear map from $\mathcal{B(P)}$ to
$\mathcal{B}(\overline{\mathcal{P}})$.

\emph{(iv)} Projecting $\mathcal{B}$ onto the support $\mathcal{P}$ of
$\Sigma$ yields a representation of $\mathcal{A}$.

\label{thmFixedAlgebra}
\end{theorem}

The proof is deferred to \cite{RBK07c}.  The central result -- that
the fixed states are isometric to a matrix algebra -- is already
implied by the fact they form a preserved code.  Notice that if
$\mathcal{E}$ is unital, $\Sigma$ \emph{coincides} with $\mathcal{A}$
-- the non-negative, trace-1 operators in $\Sigma$ directly determine
the process' optimal noiseless codes, hence its noiseless IPS.

Familiar examples of noiseless IPS correspond to specific ways in
which information is encoded in one or more blocks of $\mathcal{M}_d$
via Eq.~(\ref{structure}).  The simplest IPS corresponds to encoding
purely classical information by a choice among multiple blocks.  For a
pointer basis, in particular, all blocks are one-dimensional. Quantum
information is preserved within a single higher-dimensional block.  A
DFS is represented by a single block with a trivial co-factor, and a
NS by a single block tensored with an identity (``noise-full'')
subsystem.  The most general IPS, a \emph{hybrid quantum memory}
\cite{KuperbergIEEE03}, has $n$ blocks of (possibly) different sizes
$d_k$.  It can be concisely described by its \emph{shape}, the vector
$\{d_1,d_2,\ldots d_n\}$.

In each of the examples above, ${\cal E}$ must (by Theorem
\ref{thmNoiselessFixed}) have a set of fixed points.  For a pointer
basis, the projectors onto each PS are fixed. For a DFS, every state
on the subspace is fixed.  The fixed points associated with a NS are
less obvious.  If $\mathcal{E}$ has a NS, ${\cal H}$ may be decomposed
as $\mathcal{H} = \mathcal{H}_A \otimes \mathcal{H}_B \oplus
\mathcal{H}_C$, and for all $\rho_A$ and $\rho_B$,
$\mathcal{E}(\rho_A\otimes\rho_B) = \rho_A\otimes\sigma_B$~\cite{IF}.
That is, $\mathcal{E}$ acts on $\mathcal{H}_A \otimes \mathcal{H}_B$
as $\mathcal{E} = \Id_A \otimes\mathcal{E}_B$, and by Schauder's fixed
point theorem \cite{GranasBook03}, $\mathcal{E}_B$ must have a fixed
point $\tau_B$.  Thus, for any $\rho_A$, $\rho_A\otimes\tau_B$ is in
$\Sigma$. Note how, for each $\sigma_B$, there is a distinct noiseless
code $\mathcal{C}_\sigma = \{\rho_A\otimes\sigma_B\ \forall \rho_A\}$,
which is isometric to the unique fixed code $\mathcal{C} =
\{\rho_A\otimes\tau_B\ \forall \rho_A\}$.

In general, the explicit form of the fixed states given in Theorem
\ref{thmFixedAlgebra}(ii) illustrates what it means to be ``isometric
to a matrix algebra'': The ``noise-full'' subsystems are represented,
not by $\Id_{n_k}$, but by a fixed state $\tau_{n_k}$.  Fixed
observables have a different structure, also derived from that of
$\mathcal{A}$.  Their restriction to $\mathcal{P}$ coincides with
$\mathcal{A}$, but each has an ``echo'' of itself on
$\overline{\mathcal{P}}$.  $\mathcal{E}^\dagger$ extends observables
on $\mathcal{P}$ to $\mathcal{\overline{P}}$, so that they detect
states initially outside of $\mathcal{P}$.  This is the
Heisenberg-picture manifestation of the fact that $\mathcal{E}$ maps
states on $\overline{\mathcal{P}}$ to $\mathcal{P}$.

\textbf{Finding the Noiseless IPS:} 
By construction, $\mathcal{E}$'s noiseless IPS contains all of
$\mathcal{E}$'s NS.  To \emph{find} this IPS:

1. Write $\mathcal{E}$ as a $d^2\times d^2$ matrix.

2. Diagonalize it, and extract the $\lambda = 1$ right and left
eigenspaces ($\Sigma$ and $\mathcal{B}$, respectively).

3. Compute $\mathcal{P}$, the joint support of all $\rho\in\Sigma$,
and project $\mathcal{B}$ onto $\mathcal{P}$ to obtain a basis for
$\mathcal{A}$.

4. Find the shape of $\mathcal{A}$, using (e.g.) tools in
\cite{HolbrookQIP04}.

Our algorithm runs in time $O(d^6)$ (matrix diagonalization is
$O((d^2)^3)$), and uses standard numerical tools.  As such, it is more
efficient than algorithms (e.g., \cite{ZurekPTP93,ChoiPRL06}) that
require exhaustive search over states or subspaces in $\mathcal{H}$ --
for these sets grow exponentially in volume with $d$.

The above algorithm may be easily generalized to \emph{unitarily
noiseless} IPS, provided that we shift our focus from $\mathcal{E}$'s
fixed points to its \emph{rotating points}, defined as follows:
\emph{The rotating points of $\mathcal{E}$ comprise the span of its
unit-modulus eigenoperators.}  We then have:

\begin{theorem} Every optimal unitarily noiseless code for $\mathcal{E}$ is
isometric to the [positive trace-1 states in the] rotating points of
$\mathcal{E}$. \label{thmUnitarilyNoiseless}
\end{theorem}

The key observation for the proof (deferred to \cite{RBK07c}) is that
there exist high powers of $\mathcal{E}$ that project onto its
rotating points. Thus, $\mathcal{E}$ has a unique unitarily noiseless
IPS, which can be found using the algorithm above provided that ``the
$\lambda=1$ eigenspace'' is replaced with ``the span of all the
unit-modulus ($\lambda=e^{i\phi}$) eigenoperators''.

\textbf{Discussion:} Our IPS framework can be used in multiple ways.
An experimentalist who has characterized a system using quantum
process tomography can apply our algorithm to find noiseless and
unitarily NS -- then use the IPS shape as a concise language to report
the results.  On a theoretical front, we have classified all optimal
preserved codes.  This rules out certain kinds of information, as
unphysical -- e.g., no process acting on a single qubit can perfectly
preserve only $\Id$, $\sigma_x$, and $\sigma_y$ (a ``rebit'').

Physically, the IPS shape distills the invariant properties of a
process (\emph{what kind} of information is preserved), discarding the
details (\emph{which} states are preserved) that are needed to design
quantum hardware, but not to understand what it can do.  It is closely
related to $\mathcal{E}$'s eigenvalues, but is both more concise and
more informative~\cite{observation}. One might hope to generalize our
algorithm to find \emph{all} correctable codes, not just noiseless
ones.  However, a constructive algorithm seems difficult, and finding
the best codes for even a \emph{classical} process is NP-hard.  Thus,
while we now know what every code must look like, \emph{finding} one
may be intractable.

\textbf{Acknowledgments:} We thank H. Barnum, P. Hayden, M.-B. Ruskai,
R. Spekkens, J. Yard, P. Zanardi, and W. Zurek for helpful
discussions.  This work was supported in part by the Gordon and Betty
Moore Foundation, by the NSF under Grants No. PHY-0456720 and
PHY-0555417, and by NSERC.

\vspace*{-3mm}

\bibliographystyle{apsrev}

\begin{thebibliography}{25}
\bibitem{iso} {An isometry is a 1:1 distance-preserving map; the
relevant distance here is the 1-norm, $||A||_1=\mbox{tr}|A|$.}
\bibitem{observation} {For example, processes that preserve a 1-qubit
DFS and a 4-state pointer basis have identical eigenvalues
($\{1,1,1,1,x_1,x_2,\ldots\}$, $|x_j|<1$), but different IPS shape
(respectively, $\{2\}$ and $\{1,1,1,1\}$).}
\end{thebibliography}

\end{document}